\documentclass{elsart}
\usepackage{epsfig}

\begin{document}

\begin{frontmatter}

\title{Length, Protein-Protein Interactions, and Complexity}

\author{Taison Tan$^1$, 
Daan Frenkel$^2$, 
Vishal Gupta$^1$, and
Michael W. Deem$^{1,3}$
}

\address{
\hbox{}$^1$Department of Bioengineering and
\hbox{}$^3$Department of Physics \& Astronomy
       Rice University, Houston, TX 77005--1892\\
\hbox{}$^2$FOM Institute for Atomic and Molecular Physics
Kruislaan 407, 1098 SJ Amsterdam, The Netherlands
}

\maketitle

\begin{abstract}
The evolutionary reason for the increase in gene length from
archaea to prokaryotes to eukaryotes observed in large scale genome
sequencing efforts has been unclear.  We propose here that
the increasing complexity of protein-protein interactions has
driven the selection of longer proteins, as longer proteins
are more able to distinguish among a larger number of distinct
interactions due to their greater average surface area.
Annotated protein sequences available from the SWISS-PROT database
were analyzed for thirteen eukaryotes, eight bacteria, and two
archaea species.  The number of subcellular locations to which each
protein is associated is used as a measure of the number of
interactions to which a protein participates.  Two databases of
yeast protein-protein interactions were used as another
measure of the number of interactions to which each
\emph{S. cerevisiae} protein participates.
Protein length is shown to correlate with both
number of subcellular locations to which a protein is associated
and number of interactions as measured by yeast two-hybrid experiments.
Protein length is also shown to correlate with the probability that
the protein is encoded by an essential gene.
Interestingly, average protein length and number of subcellular locations are
not significantly different between all human proteins and
protein targets of known, marketed drugs.
Increased protein length appears to be a significant mechanism by which
the increasing complexity of protein-protein interaction networks is
accommodated within the natural evolution of species.
Consideration of protein
length may be a valuable tool in drug design, one that predicts
different strategies for inhibiting interactions in aberrant and
normal pathways.

\end{abstract}
\end{frontmatter}

\section{Introduction}

It has been noticed that genomes from the various domains of life
differ greatly in size. For example, the human genome is roughly 30 times
larger than that of the \emph{Drosophila}.
It has also been noticed that protein lengths vary systematically
within the three domains of life~\cite{Zhang}.
The eukaryote domain is found to have
the longest average protein length per genome, while
the archaea domain is found to have the shortest average protein length per
genome.  Investigations to date have not been able to pinpoint the
biological significance or evolutionary mechanism for these observations.

Following a suggestion in \cite{Zhang},
we here propose that protein lengths are correlated with genome
size in order to cope with the increased complexity of
protein-protein interactions that arises within larger genomes.
Examples of increased complexity and functionality associated with
larger systems abound in engineering and biology. In the
highly-optimized tolerance theory of Doyle, complexity of an
engineering or biological system is postulated to be a requirement
of robustness~\cite{Doyle}. It is implicitly assumed in this
theory that selection for increased robustness leads to larger,
and more complex, systems. In simple liquid crystals,
incorporation of additional molecular species can lead to the
formation of new phases~\cite{Prost}.
In materials science, it is well-known that the number of possible
material phases increases dramatically with the number of elements
present~\cite{Rhines}. In protein structural biology, it is known
that a finite number of distinct amino acids, more than two and
typically on the order of five, is required to reproduce
characteristic protein structures~\cite{Riddle}. In polymer
physics, the complexity of material phases increases greatly as
consideration is expanded from homopolymers to diblock
copolymers~\cite{Bates} to triblock polymers~\cite{Bates2}.
Organization of the collective dynamics of social networks, the
internet, or traffic flow into so-called small world networks
allows for efficient communication within large
systems~\cite{Watts}.
The latter finding is relevant in the present context, because
interactions between proteins have been shown to increase with the
number of proteins present at a rate that is consistent with
small-world-network theory~\cite{Mott}.

In this paper, we show that a positive correlation exists between protein
length and the number of subcellular locations in which a protein is
found.  For yeast, we shown additionally a positive correlation between
protein length and number of observed protein-protein interactions.
This positive correlation of complexity with length supports the hypothesis
that an increase in protein length is necessary
for an increased number of specific interactions, on average.
The positive correlation is shown to exist for prokaryotes and eukaryotes.
Based upon these considerations of length,
it is shown that most single-drug protein inhibitors
are inhibiting a particular receptor or target site on a protein,
rather than knocking out the whole protein.

\section*{Methods and Results}

Protein length comparisons across the three domains of life were performed
utilizing the Meta-A annotation of the
SWISS-PROT database \cite{Eisenhaber}.
This allowed determination of lengths of expressed proteins.
 The average protein lengths
were determined for each organism whose genome had a large number of
completely sequenced expressed proteins.
The standard error estimate
of the average was also calculated.
Thirteen eukaryotes, eight bacteria, and two archaea were
analyzed for comparisons of protein length for the entire genome.
These species were chosen because they have the highest
frequency of entries in the SWISS-PROT database.
The Meta-A database was made non-redundant by
removing entries with different accession numbers but
identical sequences as returned by a SWISS-PROT FASTA query
\cite{swissprot}, of which there were very few.
Accession numbers that corresponded to protein fragments were also
removed.
Shown in table 1 are the numbers of proteins used within each
species.

The Meta-A annotation
determines with which subcellular locations each protein is associated.
This number of subcellular locations is used
as one measure of the number of interactions in which a protein participates.
Thus, the data from the analysis of the SWISS-PROT database is separated
according to each organism and according to the domain of life,
archaea, bacteria, or eukaryote, to which the
organism belongs.  Each protein of the organism is then categorized
according to how many subcellular locations to which it is associated. The
possible subcellular locations are intracellular, membrane related,
extracellular, cytoplasmic, transmembrane, mitochondrial, chloroplast,
nuclear, endoplasmic reticulum/Golgi apparatus, viral, and DNA
binding. For the archaea domain the number of locations
ranges from one to four, for the bacteria domain the number ranges from
one to six, and for the eukaryote domain the number ranges from one to
seven. The locations nuclear, mitochondrial, chloroplast and ER/Golgi are
available only for the eukaryote proteins.  There is no protein that is
associated with all of the compartments for bacteria and eukaryotes.
The number of amino acids in each protein is downloaded
from SWISS-PROT. The average length of a protein is 
correlated  by a linear least squares fit
with the number of subcellular locations for each species
(a non-linear correlation does not substantially better represent the data).
The standard error estimate of the
slope of the correlation (and $t$-statistic) was calculated.
The number of domains contained within each protein
was also calculated, using the Pfam database (swisspfam datafile)
\cite{Pfam}, and correlated by a linear least squares fit
with the number of subcellular locations.
The standard error estimate of the
slope of the correlation (and $t$-statistic) was calculated.

Number of protein interactions for \emph{S.\ cerevisiae} is
determined from two databases, the updated data of
Uetz \emph{et al.}\ \cite{pathcalling} 
and the comprehensive MIPS database  \cite{mips}.
The number of protein-protein interactions was correlated
with protein length by a linear least squares fit,
and the standard error estimate of the
slope of the correlation (and $t$-statistic) was calculated.
It was necessary to make the set of associated interacting proteins
non-redundant for each protein within MIPS.
For both Uetz \emph{et al.}\ and MIPS, all proteins with greater than zero
interactions were used, starting from those within the Meta-A dataset.
These data were used to correlate protein length directly with number of
interactions.  Those proteins within the Meta-A protein dataset 
that are essential were determined from the
\emph{Saccharomyces} Genome Deletion Project
\cite{knockout}.  The relationship between protein length
and the probability that the encoding gene is essential 
was determined.  The standard error estimate of the probability
was also calculated.

Finally, a list of protein targets of known, marketed drugs
was constructed by selecting from the Harvard Small Molecule Bioactives
Database \cite{Harvard} those compounds that targeted a specific
protein which could be identified in SWISS-PROT. Of the greater
than 2000 compounds in the database, 186 targeted a specific protein, of which 
100 were unique.
The number of drug targets is in agreement with
Pfizer's proprietary list of protein drug targets, which
contains 120 unique proteins \cite{druggable}.

Figure 1 shows the plot of the average protein lengths
for the various genomes across the three domains of life. The mean values
of the protein lengths decrease from eukaryote to bacteria to archaea.
This trend in protein lengths is identical to that previously observed
for gene lengths \cite{Zhang}.

Figure 2 shows a positive correlation between average
protein length and number of associated subcellular locations for
the bacteria domain.
Figure 3 shows a positive correlation between average protein length
and number of associated subcellular locations for the eukaryote domain.
Positive correlations are found
for all thirteen eukaryote species shown in figure 3. A
similar correlation was not found for the archaea domain (data
not shown).
Also shown in figures 2 and 3 are the correlations between
number of domains within a protein and number of associated subcellular
locations.

The probability that a protein of a given length will be encoded by
an essential gene is shown  in figure 4.  A positive
correlation is found between length and essentiality.

The number of interactions for yeast proteins of various lengths
is shown in figure 5.  Since the data become sparse
for large numbers of interactions, only those proteins with fewer
than a dozen interactions are shown.

The average lengths and number of subcellar locations of human
proteins are shown in table 2, along with these
same quantities for the protein targets of marketed drugs.

\section*{Discussion}

That proteins must be longer in order to achieve and distinguish among
an increased number of interactions is a
mechanistic explanation for the positive correlations found in 
bacteria and eukaryotes. At a more coarse-grained level,
comparisons between human, fly, worm, and yeast have shown that
the human proteome set contains 1.8 times as many protein domains
as the worm or fly and 5.8 times as many as yeast~\cite{Lander}.  Moreover,
the average eukaryote gene length is roughly 1.4 times the
average bacterial gene length and 1.6 times the average
archaea gene length \cite{Zhang}. 
The increase in the number of domain architectures and genome complexity
correlates with the
observed increase in protein length. The larger vocabulary of
protein domain architectures allows the more evolved systems to accommodate an
increased number of interactions.

The results in figures 2 and 3 are statistically
significant.  Most significant is that if one assumed the correlations
were random, with a slope symmetrically distributed about zero, the
probability that 20 or more out of 21 of these correlations would
be positive is $\left[
\left( \hbox{}_{21}^{21} \right)
+
\left( \hbox{}_{20}^{21} \right)
 \right]2^{-21} = 10^{-5}$.
Taken in aggregate, therefore, figures 2a and 3a
suggest that there is a highly significant and positive correlation
between protein length and number of subcellular locations to
which a protein  is associated.  At the level of each individual
correlation, the 
average $t$-statistic for the positive correlations in
figure 2a is 1.80, and
that of figure 2b is 0.87.
The average $R$ values are 0.64 and 0.44, respectively.
The average $t$-statistic for the positive correlations in
Figure 3a is 2.46, and
that of figure 3b is 2.70.
The average $R$ values are 0.80 and 0.65, respectively.
Interestingly, the correlations for the eukaryotes are somewhat more
significant ($P = 0.014$ for figure 3a)
than those for the bacteria
($P = 0.072$ for figure 2a).
For the bacteria, the correlations with protein length observed
in figure 2a are slightly 
more significant than the 
the correlations with number of domains observed in 
figures 2b.  For the eukaryotes, the
correlations with protein length are of roughly the
same significance as those with number of domains.

An alternative means of correlating length with number of locations
would be to use all of the protein entries, rather than preaverage them
within each location, as in figures 2a and 3a.
When this is done for the bacteria, the positive correlations
persist, except for \emph{Aquifex aeolicus}, \emph{Bacillus substilis},
and \emph{Synechocystis sp.}
Similarly, when this is done for the eukaryotes, the positive correlations
persist, except for \emph{Saccharomyces cerevisiae}.
Overall, then, the trend of increasing number of interactions with
increasing length among the prokaryotes and eukaryotes is rather
robust to the means of measurement.
The average $t$-statistic for the positive correlations 
determined by this method for the analog of
Figure 2a is 2.5035,
and the average $t$-statistic for the positive correlations
determined by this method for the analog of
figure 3a is 6.454, with the increased
significance due to the greater number of data points.

The topological scaling properties of metabolic 
networks of organisms show similarities to complex
systems in general~\cite{Jeong}. The topology of these scale-free networks
is dominated by highly-connected nodes, ``hubs,'' that link together
the less-connected nodes. Deletion of these
hubs is an especially efficient way to destroy the connectivity of
the network~\cite{Jeong2000}. From our results, the hub proteins of
large protein-protein networks are more likely to be the longer
proteins, since we have shown that longer lengths typically
possess more inter-protein interactions. 
Figure 4, moreover, explicitly shows that longer proteins
are more essential, precisely because they are more connected
\cite{Jeong2001}.
%It would be interesting to study the
%correlation between length of
%protein and the degree to which a protein serves as a hub in a
%protein network.  

As shown in table 2, protein targets of known drugs
are not substantially shorter or less numerous in number of
subcellular interactions than are average human proteins.  Given the
above arguments about protein-protein networks, this lack of
difference might seem surprising \cite{Hasty2001}.  What this
result implies, in fact, is that most known single-drug protein inhibitors
are inhibiting a particular receptor or target site on a protein,
rather than knocking out the whole protein.

If it is desired to
disrupt a pathway with many distinct interactions, most likely the
best proteins to knock out will be the longer ones.  Therefore, it
may be advisable for multiple drug regimen therapies that target
aberrant pathways to target multiple receptors of
the longest proteins of those
pathways, within the set of otherwise equally suitable targets.
However, if minimal disruption of a normal pathway is desired, the
focus of the therapy should be on the smaller proteins or
single-drug therapies. These
smaller, shorter proteins are less likely to be at the center of
network pathway hubs, and their deletion would be least likely to
disrupt the network, all other factors being equal. From an
evolutionary point of view, shorter proteins with fewer
interactions would more readily be independently evolved, having
fewer epistatic interactions \cite{epistatic}. Directed
pathway evolution studies may benefit from a focus on such
proteins~\cite{Stemmer2002}.

It might be argued that number of subcellular locations is only
an approximate measure of the number of interactions to which a
protein participates.  For yeast, there are explicit measurements of
the number of protein-protein interactions.
Shown in figure 5 is the
positive correlation between length and number of
interactions determined in yeast when all of the
proteins with entries in SWISS-PROT are correlated.
The correlation in figure 5a is significant to the
level $P = 0.05$ ($t$-statistic $ = 1.95$),
and the correlation in figure 5b is significant to the
level $P = 6 \times 10^{-8}$ ($t$-statistic $ = 5.425$).
The R values are 0.24 and 0.32, respectively.
The positive correlation in figure 5 is in accord with
the positive correlation found for yeast in figure
3a, and this agreement provides additional justification for the
use of number of subcellular locations as a surrogate measure of
number of protein interactions.
There are, of course, exceptions to the general correlation between
protein length and number of interactions.  For example, a large number of
proteins can interact with a specific short protein if all the
interactions occur through the same binding site.   This appears
to be the case for several single-domain  RNA- and DNA-binding
proteins (such as SOH1, LSM2, RPB9,  LSM5, and SR14).  The single-domain
GTP-binding protein (TEM1) as well as the 
membrane proton channel (AT14) of the ATPase
complex are also short and bind many proteins.
Notwithstanding these and other exceptions,
the correlations in figure 5 are striking, and the
positive correlation persists even when data for all numbers of
interactions are considered (data not shown).

The positive correlations observed in the bacteria and eukaryote domains
were not observed in the archaea domain.
The correlations for the archaea remain negative also when
all proteins are used in the correlation.
Archaea have substantially fewer subcellular locations and
total numbers of proteins.  It is also evident from figure 1
that the archaea domain possesses the shortest proteins. The
archaea domain, being the least evolved, 
may, therefore, lack the necessary diversity of protein-protein
interactions necessary to drive the evolution of the increased protein lengths
observed in the bacteria and eukaryote domain.

\section*{Conclusions}
Increased protein length appears to be a significant mechanism by which
the increasing complexity of protein-protein interaction networks is
accommodated within the natural evolution of species.  It would be
interesting to measure within experimental evolution protocols
\cite{Stemmer2002} the degree to which selection for an
increased number of specific interactions within various pathways
or subsystems is a major driver of increased protein length.
It also appears that consideration of protein
length may be a valuable tool in drug design, one that predicts
different strategies for inhibiting interactions in aberrant and
normal pathways.

%\section*{Authors' contributions}
%TT and VG carried out the bioinformatics studies.  
%DF and MWD conceived of the study and participated in its
%design and coordination.  All authors read and approved the final
%manuscript.

\section*{Acknowledgments}
This research was supported by the U.S. National Institutes of
Health.

\bibliography{length}

\newpage
\clearpage

\begin{center}
{\bf Figures}
\end{center}

\begin{figure}[htbp]
\begin{center}
\epsfig{file=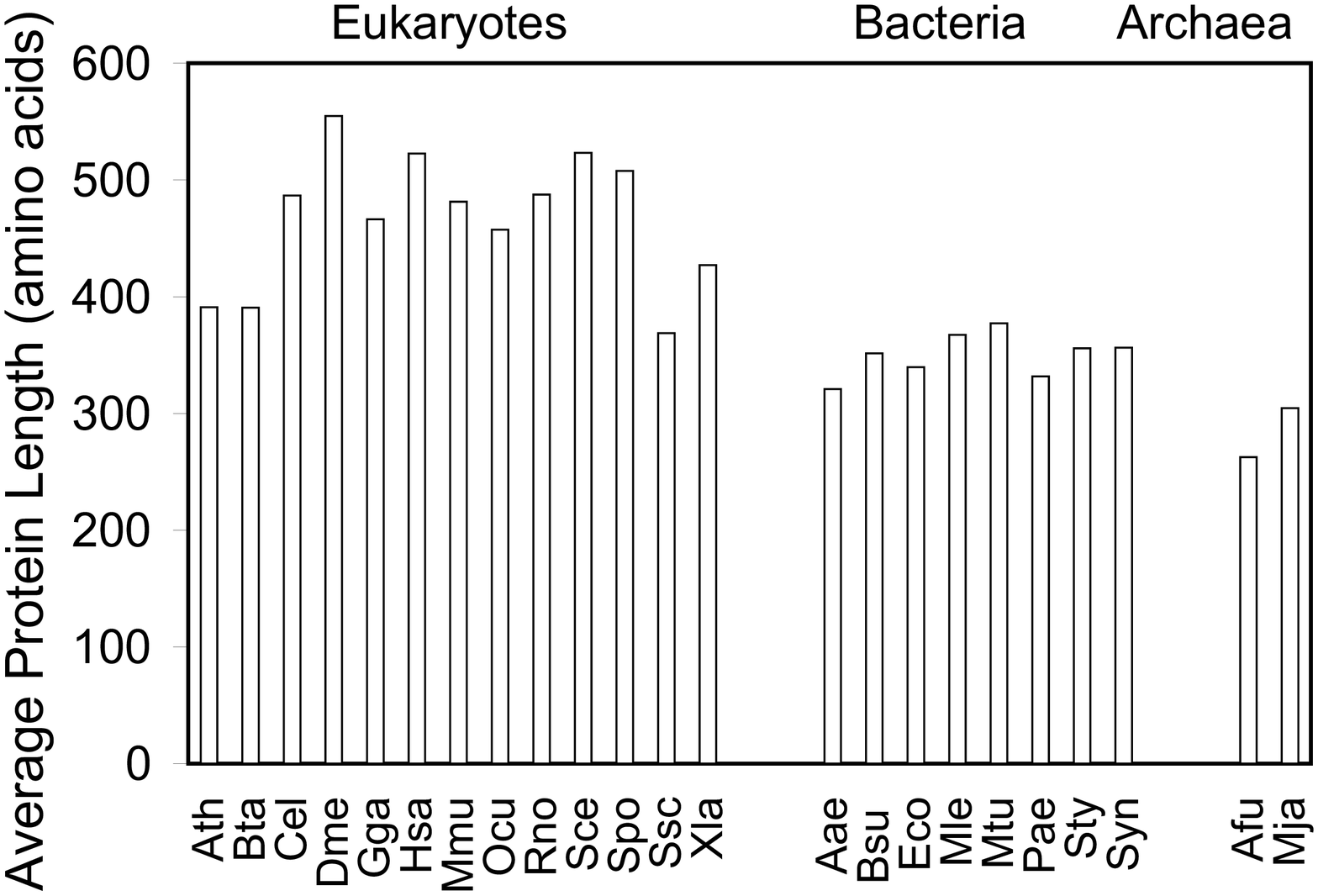,clip=,height=2in,angle=0}
\end{center}
\caption{
Mean protein length for the various genomes of the
eukaryote, bacteria, and archaea domain.
The eukaryotes studied were
\emph{Arabidopsis thaliana} (Ath),
\emph{Bos taurus} (Bta),
\emph{Caenorhabditis elegans} (Cel),
\emph{Drosophila melanogaster} (Dme),
\emph{Gallus gallus} (Gga),
\emph{Homo sapien} (Hsa),
\emph{Mus musculus} (Mmu),
\emph{Oryctolagus cuniculus} (Ocu),
\emph{Rattus norvegicus}  (Rno),
\emph{Saccharomyces cerevisiae} (Sce),
\emph{Schizosaccharomyces pombe} (Spo),
\emph{Sus scrofa} (Ssc), and
\emph{Xenopus laevis} (Xla).
The bacteria studied were
\emph{Aquifex aeolicus} (Aae),
\emph{Bacillus subtilis} (Bsu),
\emph{Escherichia coli} (Eco),
\emph{Mycobacterium leprae} (Mle),
\emph{Mycobacterium tuberculosis} (Mtu),
\emph{Pseudomonas aeruginosa} (Pae),
\emph{Salmonella typhimurium} (Sty), and
\emph{Synechocystis sp.} (strain PCC 6803) (Syn).
The archaea studied were
\emph{Archaeoglobus fulgidus} (Afu) and
\emph{Methanococcus jannaschii} (Mja).
The smallest estimate of the error was 3.63 and the largest
was 17.64 amino acids, with an average estimate of the error of 9.02
amino acids.
After \cite{Zhang}.
}
\label{fig1}
\end{figure}

\begin{figure}[htbp]
\begin{center}
\epsfig{file=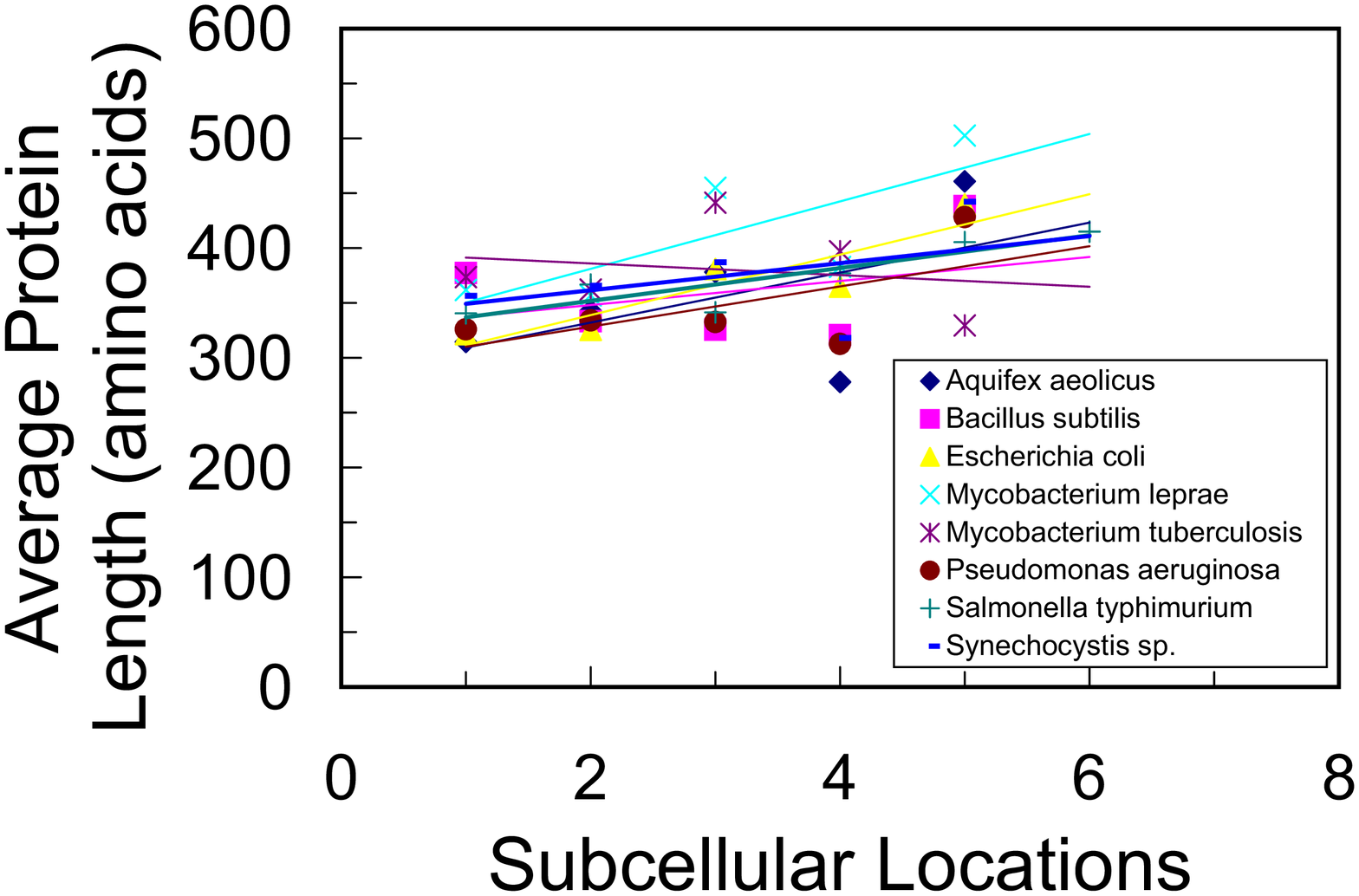,clip=,height=2in,angle=0}
\hfill
\epsfig{file=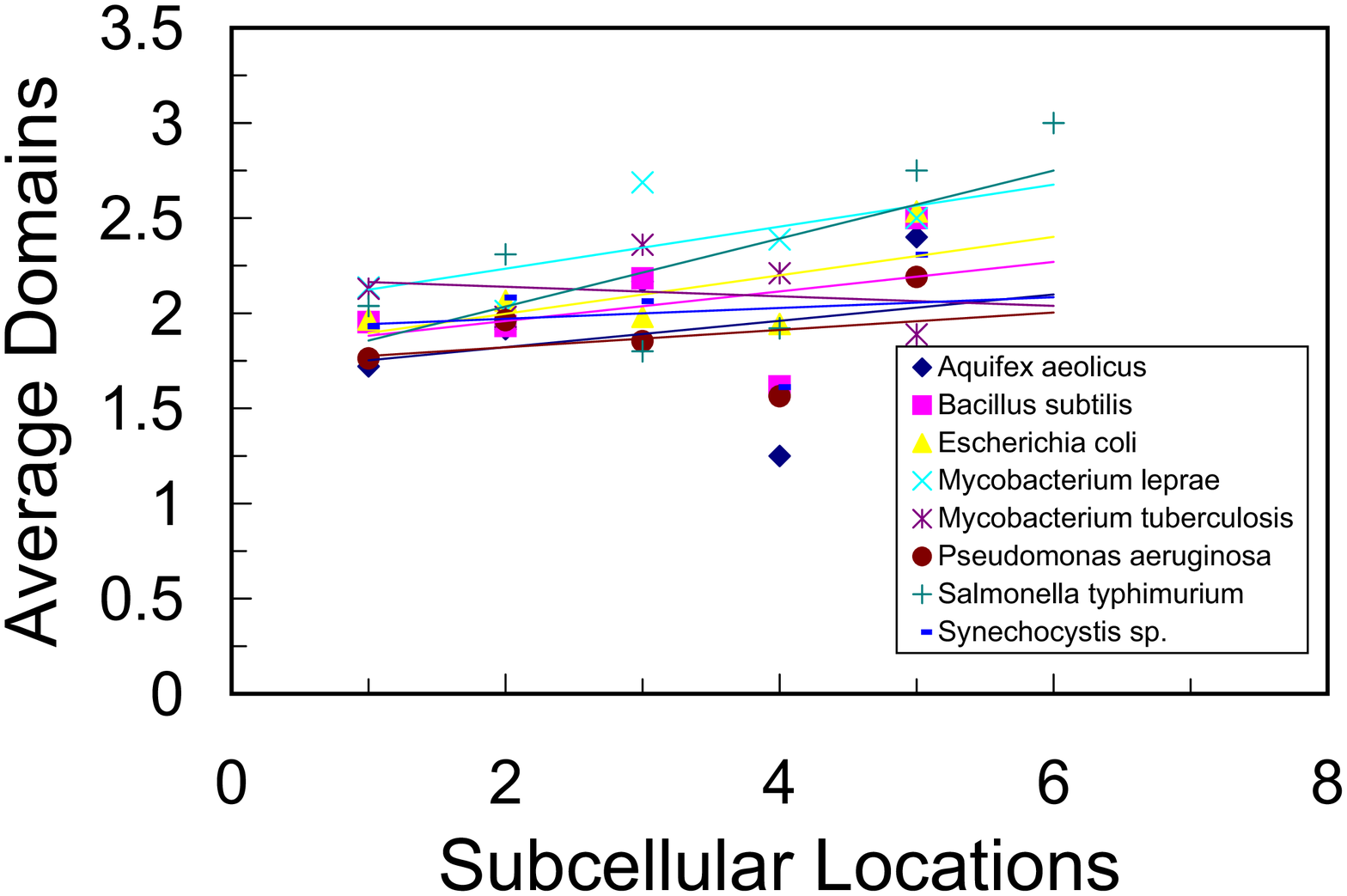,clip=,height=2in,angle=0}
\end{center}
\caption{
Correlation with number of subcellular locations
for the bacteria domain.
Positive correlation for the bacteria domain
between a) protein length or
b) number of domains within a protein and number of subcellular locations.
Only \emph{Mycobacterium tuberculosis} has a negative slope in the
correlations.
}
\label{fig2}
\end{figure}

\begin{figure}[htbp]
\begin{center}
\epsfig{file=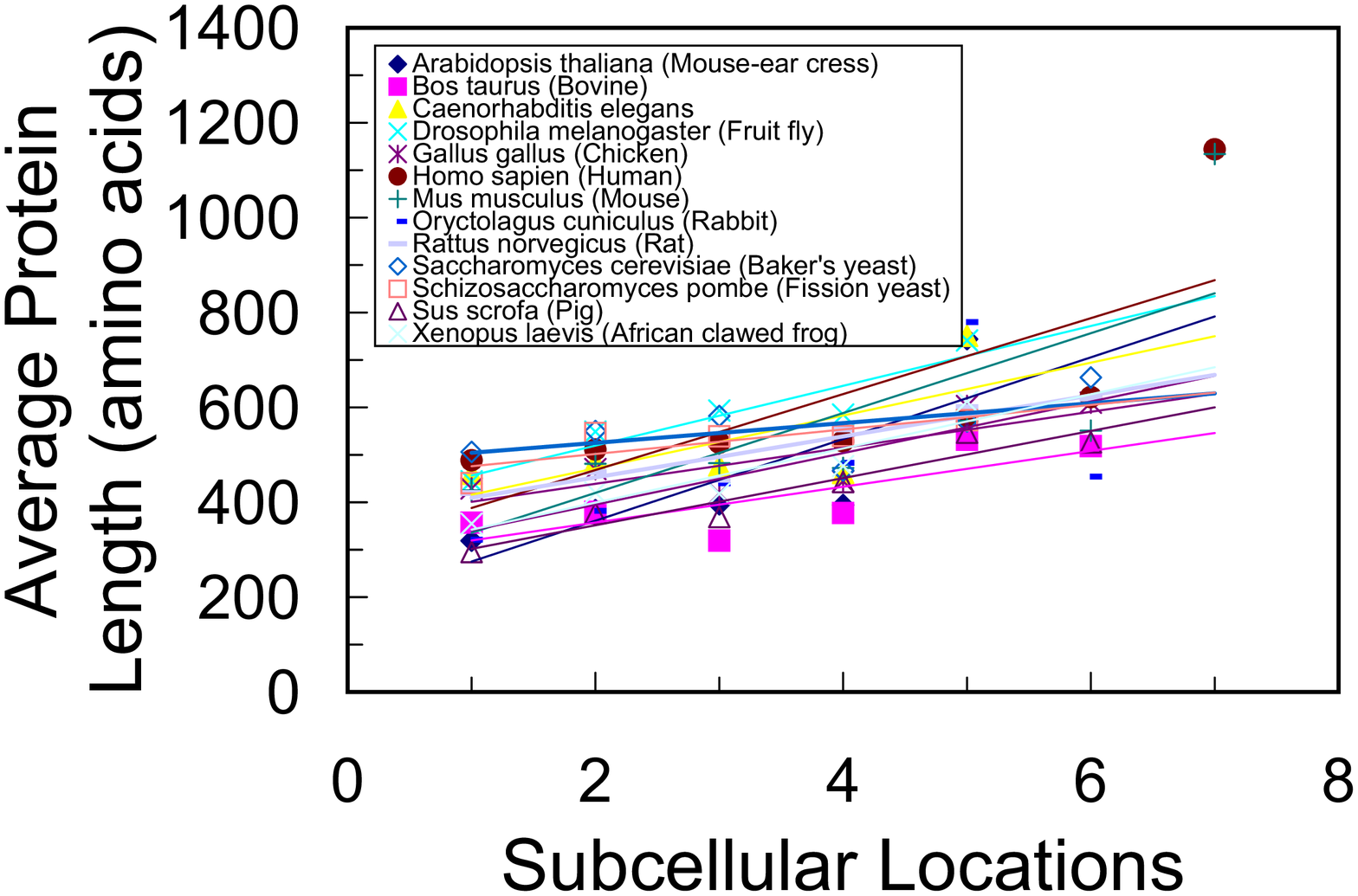,clip=,height=2in,angle=0}
\hfill
\epsfig{file=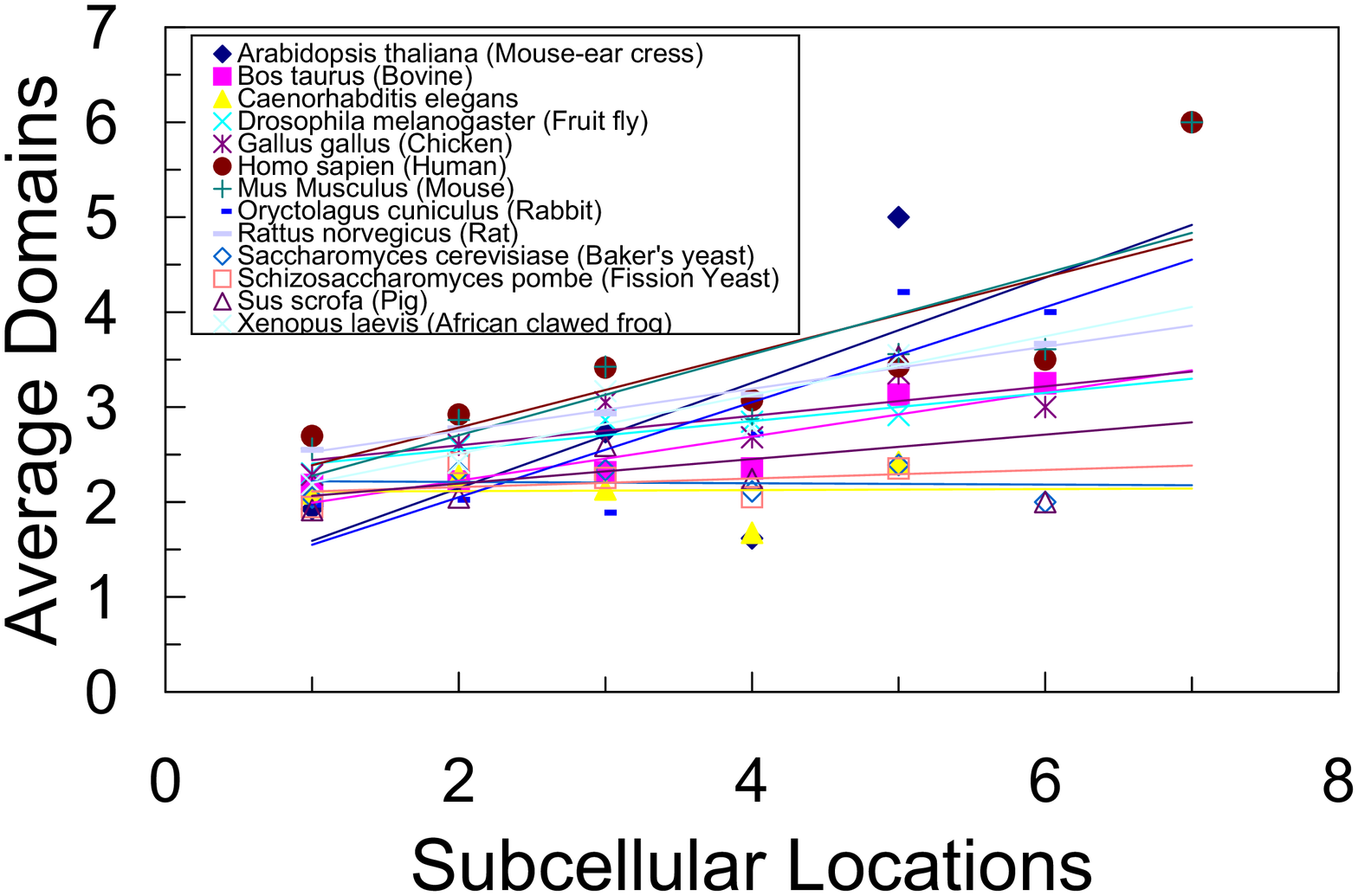,clip=,height=2in,angle=0}
\end{center}
\caption{
Correlation with number of subcellular locations
for the eukaryote domain.
Positive correlation for the eukaryote domain
between a) protein length or
b) number of domains within a protein and number of subcellular locations.
}
\label{fig3}
\end{figure}

\begin{figure}[htbp]
\begin{center}
\epsfig{file=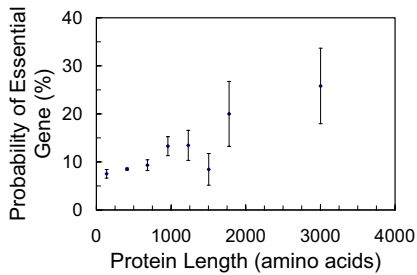,clip=,height=2in,angle=0}
\end{center}
\caption{
Correlation with essentiality for yeast proteins.
Positive correlation between the length of yeast protein and the
probability that the encoding gene is essential.  Within each
bin of protein length, the fraction of such proteins that
are essential as determined by knock-out experiments
\cite{knockout}  is plotted. 
}
\label{fig4}
\end{figure}

\begin{figure}[htbp]
\begin{center}
\epsfig{file=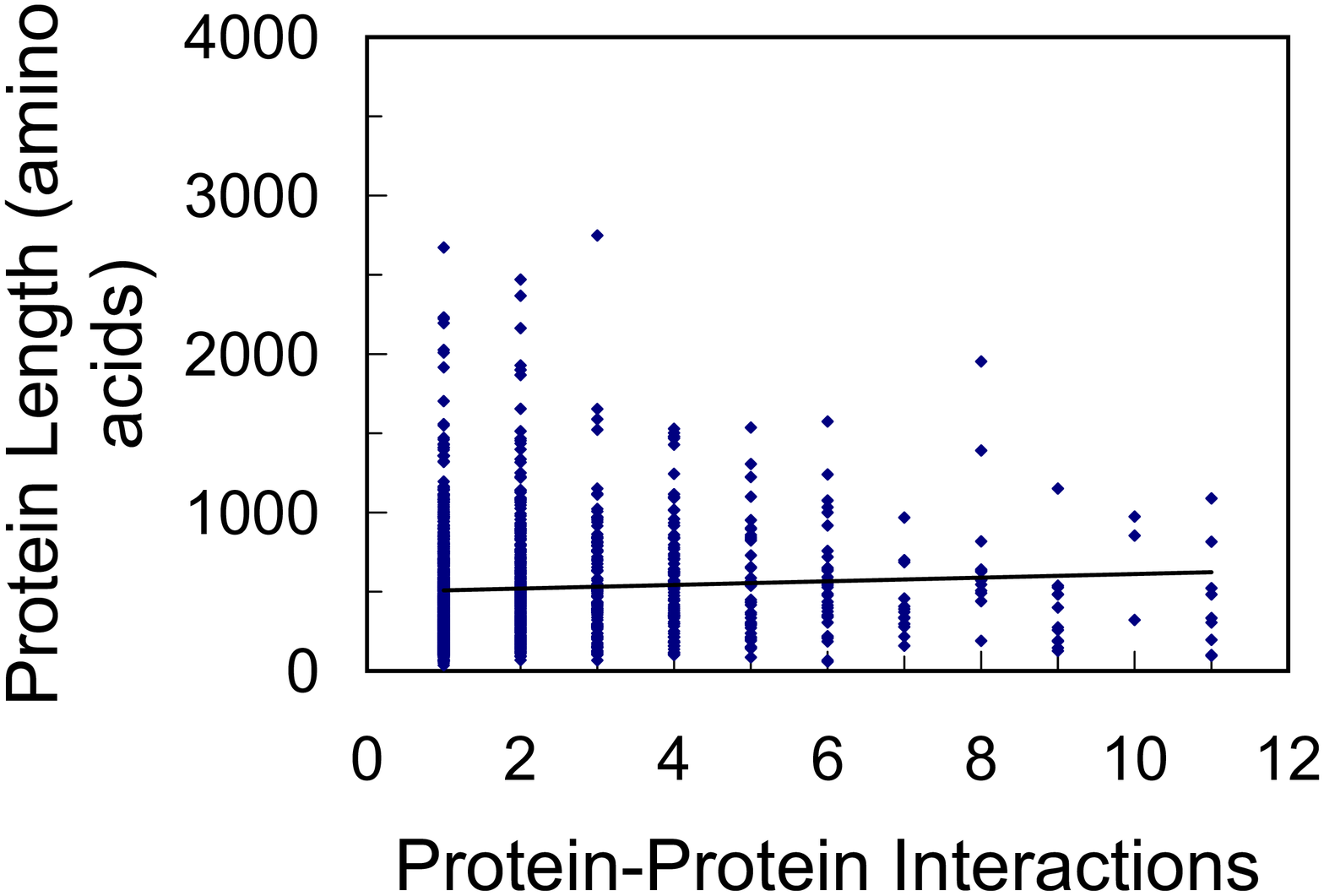,clip=,height=2in,angle=0}
\hfill
\epsfig{file=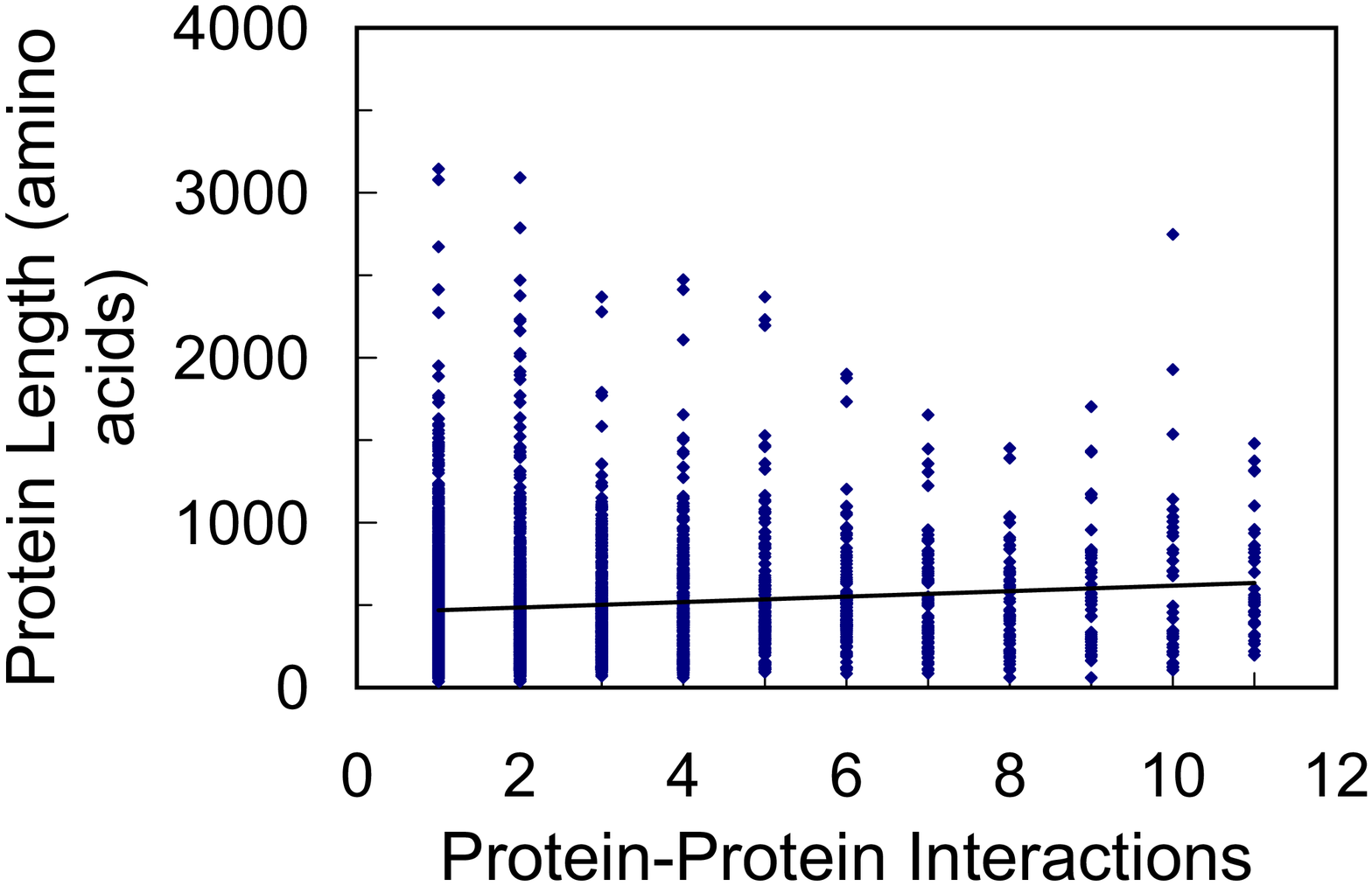,clip=,height=2in,angle=0}
\end{center}
\caption{
Correlation of protein-protein interactions in yeast.
Positive correlation for yeast between the number of
protein-protein interactions and protein length.
Data from a) \cite{pathcalling} and b) \cite{mips}.
}
\label{fig5}
\end{figure}

\clearpage
\newpage

\begin{table}[htbp]
\caption{Number of proteins annotated by Meta-A for each species.}
\label{table1}
\begin{tabular}{ll}
Species & Number of Proteins\\
\hline
Eukaryotes\\
Ath & 1313 \\
Bta & 1015  \\
Cel & 1333 \\
Dme & 1383 \\
Gga & 833 \\
Hsa & 6608 \\
Mmu & 4227 \\
Ocu & 534 \\
Rno & 2572 \\
Sce & 3394 \\
Spo & 1319 \\
Ssc & 535 \\
Xla & 648 \\
\\
Bacteria\\
Aae & 496 \\
Bsu & 1635 \\
Eco & 3478 \\
Mle & 424 \\
Mtu & 943 \\
Pae & 527 \\
Sty & 733 \\
Syn & 694 \\
\\
Archaea\\
Afu & 481\\
Mja & 769\\
\end{tabular}
\end{table}

\begin{table}[htbp]
\caption{
Length and number of subcellular locations.
Data are shown  for all human proteins and for protein drug targets
Standard error estimates are calculated from one standard deviation.
}
\label{table2}
\begin{tabular}{lllll}
~~~~~~~&\vline& All Human & Proteins Targeted & Non-redundant \\
&\vline& Proteins& by Drugs & Protein Targets\\
\hline
Number of Proteins &\vline & 6608 & 186 & 100 \\
Average Length &\vline & 523 $\pm$ 5.98 & 577 $\pm$ 22.5 & 554 $\pm$ 34.0\\
Average Number of Locations &\vline
            & 2.68 $\pm$ 0.02 & 2.64 $\pm$ 0.11 & 2.72 $\pm$  0.15 \\
Average Number of Domains &\vline & 3.06 $\pm$ 0.03 & 3.22 $\pm$ 0.14
            & 3.26 $\pm$ 0.23 \\
\end{tabular}
\end{table}

\end{document}